\documentclass[aps,prd,twocolumn,superscriptaddress,nofootinbib]{revtex4-1}
\usepackage{amsmath}
\usepackage{siunitx}
\usepackage{graphicx}
\usepackage{hyperref}
\usepackage{placeins}

\begin{document}
\title{Are ultracompact minihalos really ultracompact?}
\author{M. Sten Delos}
\email{Electronic address: delos@unc.edu}
\author{Adrienne L. Erickcek}
\email{Electronic address: erickcek@physics.unc.edu}
\affiliation{Department of Physics and Astronomy, University of North Carolina at Chapel Hill, Phillips Hall CB3255, Chapel Hill, North Carolina 27599, USA}
\author{Avery P. Bailey}
\affiliation{Department of Astrophysical Sciences, Princeton University, Peyton Hall,
Princeton, New Jersey 08544, USA}
\affiliation{Department of Physics and Astronomy, University of North Carolina at Chapel Hill, Phillips Hall CB3255, Chapel Hill, North Carolina 27599, USA}
\author{Marcelo A. Alvarez}
\affiliation{Berkeley Center for Cosmological Physics, Campbell Hall 341, University of California, Berkeley, California 94720, USA}

\begin{abstract}
Ultracompact minihalos (UCMHs) have emerged as a valuable probe of the primordial power spectrum of density fluctuations at small scales.  UCMHs are expected to form at early times in regions with ${\delta\rho/\rho \gtrsim 10^{-3}}$, and they are theorized to possess an extremely compact ${\rho\propto r^{-9/4}}$ radial density profile, which enhances their observable signatures.  Nonobservation of UCMHs can thus constrain the primordial power spectrum.  Using $N$-body simulations to study the collapse of extreme density peaks at ${z \simeq 1000}$, we show that UCMHs forming under realistic conditions do not develop the ${\rho\propto r^{-9/4}}$ profile and instead develop either ${\rho\propto r^{-3/2}}$ or ${\rho\propto r^{-1}}$ inner density profiles depending on the shape of the power spectrum.  We also demonstrate via idealized simulations that self-similarity---the absence of a scale length---is necessary to produce a halo with the ${\rho\propto r^{-9/4}}$ profile, and we argue that this implies such halos cannot form from a Gaussian primordial density field.  Prior constraints derived from UCMH nonobservation must be reworked in light of this discovery.  Although the shallower density profile reduces UCMH visibility, our findings reduce their signal by as little as $\mathcal O(10^{-2})$ while allowing later-forming halos to be considered, which suggests that new constraints could be significantly stronger.
\end{abstract}

                              
\maketitle

\section{Introduction}
The matter structure that we observe in the Universe today, appearing in such forms as galaxies and galaxy clusters, is understood to have grown by gravitational attraction from small fluctuations in the primordial matter density field.  These fluctuations also manifest themselves in the cosmic microwave background (CMB), and their properties are well understood at large scales.  In particular, primordial density fluctuations are observed to obey Gaussian statistics \cite{lewis2016planck}, allowing them to be solely described by a power spectrum $\mathcal{P}(k)$ quantifying the power contained in fluctuations at scale wave number $k$.  The primordial power spectrum is well constrained at large scales by the CMB \cite{hlozek2012atacama} and down to wavelengths as small as \SI{2}{Mpc} by the Lyman-$\alpha$ forest \cite{bird2011minimally}.  At these scales, observations are consistent with a power law ${\mathcal{P}(k)\propto k^{n_s-1}}$ with ${n_s = 0.9667\pm 0.0040}$ \cite{2016planck}.  This nearly scale-invariant spectrum is predicted \cite{lidsey1997reconstructing} by the simplest models of inflation \cite{guth1981inflationary,albrecht1982cosmology,linde1982new}.  However, numerous inflationary models yield spectra departing from scale invariance at small scales, whether through features in the inflaton potential  \cite{salopek1989designing,starobinsky1992,ivanov1994inflation,starobinsky1998beyond,joy2008new}, multiple fields \cite{silk1987double,polarski1992spectra,adams1997multiple,achucarro2011features,cespedes2012importance}, particle production \cite{chung2000probing,barnaby2009particle,barnaby2010features}, or other effects \cite{kofman1988nonflat,ashoorioon2009energy,barnaby2011large,barnaby2012gauge,randall1996supernatural,copeland1998black,martin2000nonvacuum,martin2001trans,ben2010cosmic,stewart1997flattening,covi1999running,covi1999observational,gong2011waterfall,lyth2011contribution,bugaev2011curvature}.  Noninflationary processes can also amplify small-scale fluctuations \cite{schmid1999amplification,erickcek2011reheating}.  In order to probe inflation and other early-Universe processes, it is important to extend our knowledge of $\mathcal{P}(k)$ to smaller scales.

Access to sub-Mpc scales is limited by Silk damping in the CMB and the impact of baryonic feedback on structure formation, but excessive power at such scales can still produce observable effects.  Small-scale primordial fluctuations with fractional density excess ${\delta\equiv\delta\rho/\rho \gtrsim 0.3}$ would collapse into primordial black holes (PBHs), and constraints on their abundance supply robust upper bounds on the power spectrum \cite{josan2009generalized}.  Stronger constraints are obtained from the absence of CMB spectral distortions \cite{chluba2012probing}, but the most stringent constraints come from ultracompact minihalos (UCMHs) \cite{ricotti2009new}.  Density fluctuations with amplitude ${\delta\gtrsim 10^{-3}}$, while too small to form PBHs, are still large enough to form collapsed minihalos long before the galaxy-scale structure that we observe today begins to form; UCMHs are taken to be structures that formed at redshift ${z \gtrsim 1000}$.  Importantly, velocity dispersions are small at these early times, so UCMHs are assumed to follow the most extreme form of the radial infall similarity solution \cite{fillmore1984self,bertschinger1985self} and develop a ${\rho\propto r^{-9/4}}$ radial density profile.

This profile is much more compact than that of galaxy-scale halos \cite{frenk2012dark}, a property that greatly boosts observational signatures and has led to considerable interest in UCMHs.
In the popular and well-motivated weakly interacting massive particle (WIMP) model of dark matter \cite{jungman1996supersymmetric,bergstrom2000non,bertone2005particle}, WIMPs are thermally produced in the early Universe and can therefore annihilate within UCMHs; constraints on UCMH abundance are then obtained from nonobservation of an expected annihilation signal 
\cite{scott2009gamma,josan2010gamma,yang2011constraints,yang2011abundance,saito2011primordial,bringmann2012improved,berezinsky2013formation,yang2013neutrino,yang2017tau,1712.01724} or from indirect effects \cite{zhang2011impact,yang2011new,yang2013contribution,yang2016contributions,clark2017heating}.  Constraints have also been derived for decaying dark matter models \cite{yang2013dark,zheng2014constraints,yang2014constraints}, and model-independent constraints can be obtained from UCMH gravitational lensing signatures \cite{li2012new,zackrisson2013hunting,clark2015investigatingI,*clark2016erratumI,clark2015investigatingII,*clark2017erratumII}.  These abundance constraints lead to constraints on the power spectrum and hence on inflationary models \cite{aslanyan2016ultracompact} and reheating \cite{choi2017new}.  UCMH abundance has also been used to constrain PBHs \cite{kohri2014testing} and cosmic strings \cite{anthonisen2015constraints}.

While these constraints are calculated assuming UCMHs develop the ${\rho\propto r^{-9/4}}$ profile, this profile is derived in an idealized picture satisfying spherical symmetry, radial motion, and (after halo collapse) self-similarity, or the absence of a scale length.  It has been reproduced in $N$-body simulations from carefully constructed self-similar initial conditions \cite{vogelsberger2009caustics,vogelsberger2011non} but not from more realistic conditions \cite{huss1999universal,macmillan2006universal,bellovary2008role}.  
In contrast, dark matter halos at galaxy scales form by hierarchical clustering, and simulations of this scenario yield halo density profiles of the Navarro-Frenk-White (NFW) form \cite{navarro1996structure,navarro1997universal}, which scales as ${\rho\propto r^{-1}}$ within the innermost region.  A slightly steeper profile arises in simulations with a free-streaming cutoff scale below which there are no fluctuations: halos at that scale form by steady accretion and develop ${\rho\propto r^{-3/2}}$ inner profiles \cite{ogiya2017sets,angulo2017earth,anderhalden2013density,*anderhalden2013erratum,ishiyama2010gamma,ishiyama2014hierarchical,
polisensky2015fingerprints}.  Neither of these pictures comes close to reproducing the UCMH density profile.  However, previous numerical experiments did not explore the collapse of an extreme density fluctuation at ${z\simeq 1000}$, and prior UCMH analyses have assumed that the combination of small velocity dispersion and isolation associated with a halo forming at such early times will suffice to yield the ${\rho\propto r^{-9/4}}$ profile.  A boost to the power spectrum is necessary to effect any halo formation at ${z\simeq 1000}$, but following this reasoning, the boost should be weak enough that such halos are highly isolated.

Recent simulation work in Ref.~\cite{gosenca20173d}, hereafter GABH, has called the applicability of the ${\rho\propto r^{-9/4}}$ profile into question.  GABH simulated early structure growth in a power spectrum with a narrow boost (which we will call a ``spike'') and concluded that the resulting halos have NFW profiles.  However, their analysis failed to rule out UCMHs with the ${\rho\propto r^{-9/4}}$ profile because they did not select for extreme density peaks, instead simulating a typical box whose largest peak corresponded to a $4.3\sigma$ fluctuation (smoothed at the scale of the spike).  Moreover, GABH claimed that the absence of ${\rho\propto r^{-9/4}}$ profiles owes to the lack of spherical symmetry and isolation in realistic conditions.  Since Refs.~\cite{vogelsberger2009caustics,vogelsberger2011non} observed (and we confirm) that spherical symmetry is unnecessary for the ${\rho\propto r^{-9/4}}$ profile, this claim suggests that halos forming from peaks rarer than $4.3\sigma$ could be sufficiently isolated to develop it.  Reference~\cite{bringmann2012improved} used peaks as extreme as $6\sigma$ to derive observational constraints, above the $5\sigma$ level that GABH claimed based on idealized simulations to not produce the ${\rho\propto r^{-9/4}}$ profile.

Our approach differs in that we search millions of density fields to find a sufficiently extreme peak.  We simulate $6.8\sigma$ peaks collapsing at ${z\simeq 1000}$ in a more weakly boosted power spectrum than that of GABH.  This level of statistical extremity corresponds to a UCMH mass fraction ${f\simeq 10^{-9}}$ in the analysis of Ref.~\cite{bringmann2012improved}, which is well below constrained levels\footnote{While our power spectrum spike peaks above the generalized constraint in Ref.~\cite{bringmann2012improved}, it is not ruled out because it is not locally scale invariant.}, implying that our density peak is rarer than any level hitherto assumed to suffice for development of the ${\rho\propto r^{-9/4}}$ profile.  We also consider an alternative power spectrum amplification---a step instead of a spike---and show that the density profile depends on the shape of the power spectrum.  Finally, we use idealized simulations to argue that self-similarity is necessary to produce the ${\rho\propto r^{-9/4}}$ profile, which would definitively rule out its appearance in a Gaussian random field.

\begin{figure}[t]
	\centering
	\includegraphics[width=8.5cm]{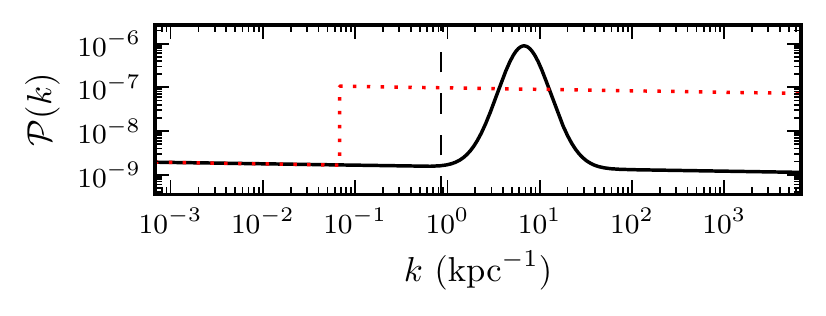}
	\caption{\label{fig:prim} Dimensionless primordial power spectrum of curvature fluctuations.  The solid line shows the spike modification, while the dotted line shows the step.  The dashed line indicates the smallest $k$ (largest scale) accessible in our simulations.}
\end{figure}

\begin{figure*}[t]
	\centering
	\includegraphics[width=\textwidth]{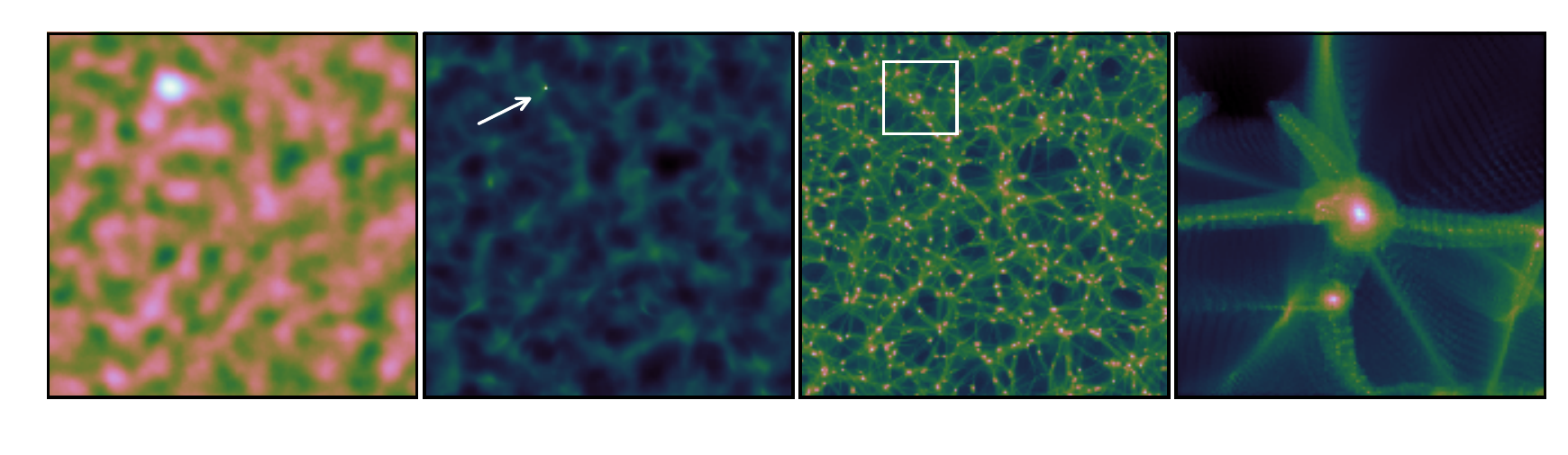}
	\caption{\label{fig:halo} $(\SI{7.4}{kpc})^3$ density field with narrowly amplified power spectrum (solid line in Fig.~\ref{fig:prim}).  Left: Thin slice of the initial density field at ${z=8\times 10^6}$.  Center left: Projected field at ${z=715}$.  The indicated halo, which collapsed at ${z\simeq 1000}$, is still the only halo visible, a testament to its rarity.  Center right: Projected density field at ${z=100}$.  The boxed region is shown in an expanded picture on the right (projected over a smaller depth).  Color scales are logarithmic; lighter indicates denser.}
\end{figure*}

\section{Simulating rare collapse}
We choose to study density fluctuations with wavelengths of order \SI{1}{kpc}.  The simulation starting redshift is set at ${z=8\times10^6}$ so that an overdense region that collapses at ${z\simeq 1000}$ initially has ${\delta\simeq 0.1}$.  
To prepare our initial conditions, we calculate a power spectrum at ${z=1000}$ using the Boltzmann code \textsc{camb sources} \cite{challinor2011linear,lewis200721} with Planck cosmological parameters \cite{2016planck} and then extrapolate it back to $z=8\times10^6$ using analytic linear theory\footnote{We use an analytic calculation that neglects baryonic effects to match simulation behavior.} \cite{hu1996small}.
We consider two types of modification to the power spectrum in order to effect the collapse of halos at ${z\simeq 1000}$.  In the first, shown as the solid line in Fig.~\ref{fig:prim}, we amplify density fluctuations over a narrow range of scales, forming a spike in the power spectrum centered at ${k=\SI{7}{kpc^{-1}}}$.  This enhancement will induce a characteristic separation between halos, allowing them to grow in isolation, a situation that best matches the canonical UCMH picture.  Our particular spike boosts the power spectrum by a peak factor of 625 and contains 90\% of its integrated power inside one $e$-fold in $k$.  The second modification, shown as the dotted line in Fig.~\ref{fig:prim}, appears as a ``step'' and represents amplification of fluctuations over a wide range of scales.  The boost factor in this case is 64, tuned to produce a similar halo number density to the spike at ${z\simeq 1000}$.  Halos in this picture will grow hierarchically, so there is no reason to expect them to differ from conventional halos, and we will only consider this case briefly.

With power spectra chosen, we search Gaussian random fields in a comoving periodic ${(\SI{7.4}{kpc})^3}$ box generated at the initial redshift.  We select a box based on the criterion that the linearly evolved field at ${z=1000}$, smoothed over ${10^{-2}\ \si{kpc}}$, has a peak with ${\delta>1.686}$, the linear threshold for collapse.  Once we have such a peak, we construct initial conditions for our simulation using the Zel'dovich approximation \cite{zel1970gravitational}.  Finally, in order to accurately model dynamics at early times, we simulate the box using a version of the TreePM code \textsc{gadget-2} \cite{springel2005cosmological,springel2001gadget} that we modified to include a smooth radiation component.\footnote{We explicitly checked that our modification reproduces the results of linear theory in the linear regime.}

We now study the results from the spiked power spectrum (solid line in Fig.~\ref{fig:prim}).  After generating over two million simulation boxes, we found nine matching our collapse criterion.  We simulated all nine boxes and found them to produce similar results, and we present a representative result here.  This simulation was carried out with $1024^3$ ($512^3$) particles to ${z=100}$ (${z=50}$), with inner structure further resolved by resimulating the first halo at $8\times$ particle density to ${z=100}$; as a result, that halo contained $4.6\times 10^6$, $1.1\times 10^7$, $1.9\times 10^7$, and $4.3\times 10^5$ simulation particles within its virial radius at ${z=400}$, ${z=200}$, ${z=100}$, and ${z=50}$ respectively.  The halo mass at these redshifts was $8.1 M_\odot$, $18 M_\odot$, $31 M_\odot$, and $51 M_\odot$.  Figure~\ref{fig:halo} shows the density field at various times.  The leftmost picture shows the initial field, emphasizing how extreme the largest peak is compared to its surroundings.  The center left picture shows that the first halo, forming near ${z= 1000}$, is still the only halo in the box by ${z=715}$.  The box at ${z=100}$ (center right) clearly displays the imprint of the spike in the power spectrum, for we see an almost uniform distribution of halos with no large-scale structure.  This is quite unlike a hierarchical clustering picture (cf. Ref.~\cite{frenk2012dark}).  There is also minimal substructure within halos, as emphasized by the rightmost picture.  Fragmentation is visible in the filaments but is not expected to affect our result.\footnote{Fragmentation is a numerical artifact present in simulations with suppressed small-scale power \cite{angulo2013warm}, but due to our convergence testing, we do not expect it to affect the density profile of the early-forming halo.  Moreover, it is also present when we reproduce the similarity solution in Fig.~\ref{fig:94}, so it cannot be the feature that prevents formation of the steep inner profile.}

\begin{figure}[t]
	\centering
	\includegraphics[width=\columnwidth]{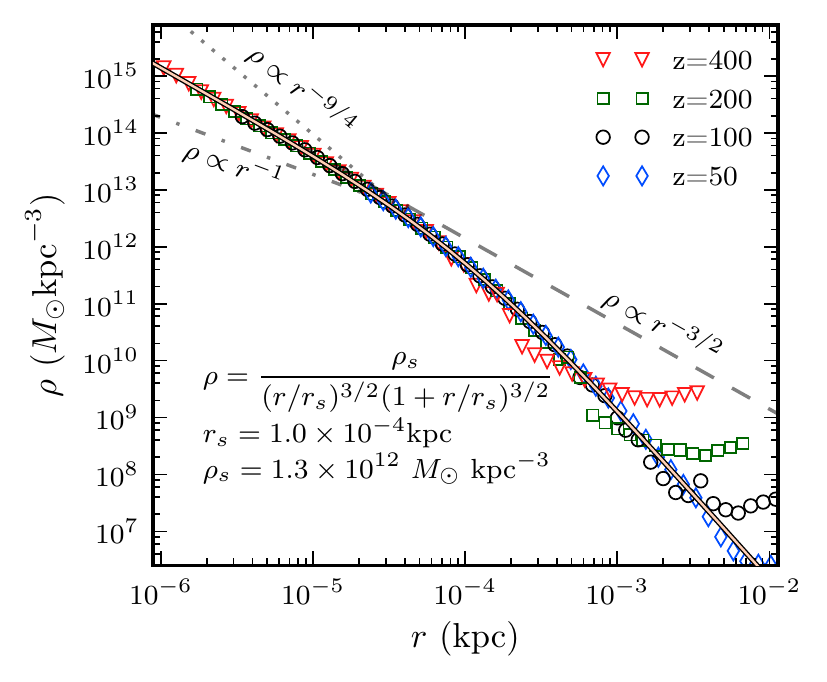}
	\caption{\label{fig:spike} Density profile in physical (not comoving) coordinates at ${z=400}$, ${z=200}$, ${z=100}$, and ${z=50}$ for a halo that forms at ${z\simeq 1000}$ due to a narrow power spectrum amplification.  The density profile approaches ${\rho\propto r^{-3/2}}$ at small $r$ and ${\rho\propto r^{-3}}$ at large $r$, following the fitting form (solid line) of Eq.~(\ref{M99}).  Several power-law curves are shown for visual reference.}
\end{figure}

We now examine the spherically averaged density profile of the first halo.  Figure~\ref{fig:spike} shows the profile at ${z=50}$, ${z=100}$, ${z=200}$, and ${z=400}$ plotted in physical (not comoving) coordinates.  We first draw attention to the dotted line, which shows a ${\rho\propto r^{-9/4}}$ density profile for comparison.  The halo clearly does not follow this form, and we have conducted extensive convergence testing to confirm the validity of this result\footnote{The density profile is stable with respect to particle count, force accuracy, force softening, and time step; see Ref.~\cite{delos2018} (forthcoming) for details.  The smallest radius plotted at each redshift contains ${N>3000}$ particles and is larger than $2.8\epsilon$, where $\epsilon$ is the softening length set at $0.03$ times the mean interparticle spacing.} for ${r>\SI{1.6e-6}{kpc}}$.  The inner profile instead scales as ${\rho\propto r^{-3/2}}$, which is still steeper than NFW (note the dot-dashed line) but matches the structure observed in simulations of the smallest halos forming above a free-streaming cutoff \cite{ogiya2017sets,angulo2017earth,anderhalden2013density,*anderhalden2013erratum,ishiyama2010gamma,ishiyama2014hierarchical,polisensky2015fingerprints}.  In retrospect, we might have expected this result, because the two pictures---the spike and the cutoff---are similar in their lack of structure below the scale of the halo.  We also see that the density profile is fixed in time, which is explained by the observation that at late times, the halo's potential well is so deep that infalling matter passes through too quickly to significantly affect the central density.  Such behavior is also exhibited by the ${\rho\propto r^{-9/4}}$ similarity solution \cite{bertschinger1985self} and by NFW halos \cite{bullock2001profiles}; it is this property that steepens the outer profile \cite{lu2006origin,zhao2003growth}.

We fit the form \cite{moore1999cold}
\begin{equation}\label{M99}
\rho(r) = \frac{\rho_s}{(r/r_s)^{3/2}(1+r/r_s)^{3/2}},
\end{equation}
the analogue of the NFW density profile for a ${\rho\propto r^{-3/2}}$ inner profile, to the density profile within the halo virial radius at all four redshifts.  Figure~\ref{fig:spike} shows that this form (solid line) provides a close fit.  Because the inner density profile does not evolve, it is plausible that the same parameters would describe the density profile today: we will take advantage of this argument later in evaluating the gamma-ray luminosity of the halo.  We remark that the scale radius ${r_s = \SI{1e-4}{kpc}}$ obeys
${r_s \simeq [(1+z_\mathrm{coll})k_\mathrm{spike}]^{-1},}$
where ${k_\mathrm{spike}= \SI{7}{kpc^{-1}}}$ was the comoving scale of the spike in the power spectrum and ${z_\mathrm{coll}\simeq 1000}$ is the halo collapse redshift.  The right-hand side of this relation is precisely the physical length scale of the spike at the time of halo collapse.  We also find
${\rho_s \simeq 30 (1+z_\mathrm{coll})^3\rho_0,}$
where $\rho_0$ is the mean matter density today; the right-hand side of this equation is proportional to the matter density at the collapse time.
We have confirmed that these scalings are approximately obeyed by later halos forming in the same simulation box, and we probe them further in forthcoming work \cite{delos2018}.

We also briefly report the results from the power spectrum with a step enhancement.  This case has density fluctuations broadly distributed across scales, so we expect structure to form by hierarchical clustering.  Indeed, the first halo in this simulation box collapsed near ${z=1000}$ as planned, but it undergoes multiple merger events between ${z=1000}$ and ${z=200}$.  The final density profile is NFW with small radii sufficiently resolved to indicate an asymptotic form at least as shallow as ${\rho\propto r^{-1}}$.  Unsurprisingly, the halos in this picture, even extreme ones collapsing by ${z=1000}$, still possess the relatively shallow inner profiles that are characteristic of late-time galaxy-scale halos.  We have now considered two different enhancements to the small-scale power spectrum, a narrow spike and a uniform amplification, and found them to produce halos with different density profiles.  In forthcoming work \cite{delos2018}, we will vary the shape of the spike to explore the transition between these two regimes.


\section{Idealized simulations}\label{sec:ideal}

\begin{figure}[t]
	\centering
	\includegraphics[width=\columnwidth]{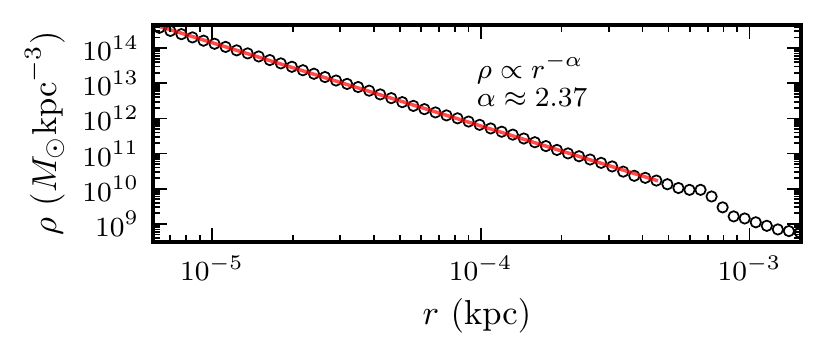}
	\caption{\label{fig:94} The density profile at ${z=200}$ of a halo that formed from a uniform ellipsoid.  This follows a pure power law ${\rho\propto r^{-\alpha}}$ (solid line) with ${\alpha > 2}$, as predicted by radial infall theory.}
\end{figure}

We now check whether we can produce the ultracompact ${\rho\propto r^{-9/4}}$ density profile from an initially uniformly overdense ellipsoid.  We construct the initial peak at ${z=\num{18000}}$ to collapse near ${z=1000}$, and we evolve it under matter domination (so radiation is neglected) to $z=200$, at which time its radial density profile is plotted in Fig.~\ref{fig:94}.  We find that the density profile is well described by a pure power law ${\rho\propto r^{-\alpha}}$ with ${\alpha > 2}$.  While the power-law index $\alpha$ is not exactly $-9/4$, the best-fit $\alpha$ is even steeper at ${\alpha\simeq 2.38}$.  We conclude that our simulation parameters are sufficient to produce such steep cusps, and that it is for physical reasons that they do not appear in realistic simulations.  This test also confirms the finding of Refs~\cite{vogelsberger2009caustics,vogelsberger2011non} that spherical symmetry is not important to the production of the ${\rho\propto r^{-9/4}}$ density profile.

Let us recount three scenarios under which the ${\rho\propto r^{-\alpha}}$ similarity solution has been reproduced in three-dimensional simulations.  The first, presented in  Refs.~\cite{vogelsberger2009caustics,vogelsberger2011non}, used carefully constructed initial conditions to match the structure of the analytic similarity solution of Ref.~\cite{fillmore1984self}.  The second, presented in GABH, begins with an initial peak in the shape of a radial Gaussian function.  The third, presented here, begins with a uniformly overdense ellipsoid.  The key feature shared between these scenarios is self-similarity, or the absence of a scale length.  This is obvious in the first case but more subtle in the other two.  A uniform ellipsoid or radial Gaussian clearly has a scale length, but only within a finite (or effectively finite) region.  This initial structure does not significantly affect the final density profile because it collapses as the halo first forms, and outside of it, the fractional mass excess contained within a given radius obeys ${\delta M/M \propto M^{-1}}$: it possesses self-similarity of the form treated in Ref.~\cite{fillmore1984self}.  This common feature between otherwise disparate initial conditions strongly suggests that self-similarity is a necessary condition to produce the ${\rho\propto r^{-9/4}}$ profile.

It is important to note that this form of self-similarity only holds for uncompensated peaks.  Peaks generated by a localized boost to the power spectrum are compensated by a surrounding trough, which maintains the scale length indefinitely.  To study how this permanent scale length affects the density profile, we employed the description of Ref.~\cite{bardeen1986statistics} to construct an idealized initial peak drawn from our spiked power spectrum of Fig.~\ref{fig:prim} with amplitude ${\delta\simeq 0.1}$ constrained in order to effect collapse at ${z\simeq 1000}$.  This peak is necessarily in isolation and possesses no substructure.  We simulated it from ${z=8\times 10^6}$ to ${z=100}$ and found that it produced the same density profile as Fig.~\ref{fig:spike}.  This is consistent with the results of Ref.~\cite{ogiya2017sets}, who carried out similarly idealized simulations of halos in a cutoff power spectrum, and we conclude that isolation alone is not sufficient to produce the ${\rho\propto r^{-9/4}}$ density profile.

While isolation is evidently insufficient to produce the ${\rho\propto r^{-9/4}}$ profile, it is also clearly necessary.  Halos that undergo major mergers develop isotropic velocity fields \cite{li2007assembly}, which produce the ${\rho\propto r^{-1}}$ inner profile characteristic of hierarchical growth \cite{lu2006origin}.  GABH also confirmed via idealized simulations that an uncompensated peak must have much greater amplitude than surrounding structure to maintain its ${\rho\propto r^{-9/4}}$ profile.


\section{Discussion and impact}
We argued in Sec.~\ref{sec:ideal} that self-similarity and isolation are both necessary conditions for the development of the ${\rho\propto r^{-9/4}}$ profile.  We now claim that these conditions cannot both be satisfied in a Gaussian density field.  Consider first the case where fluctuations are enhanced over a wide range of scales, as in our ``step'' modification.  This picture produces a nearly self-similar peak in the spherical average, but it also generates an abundance of structure at ever smaller scales.  Isolation fails here as structure grows hierarchically, resulting in ${\rho\propto r^{-1}}$ inner profiles.  We can instead preserve isolation by suppressing small-scale power by creating a spike (or cutoff) in the power spectrum, but this feature imposes a scale length, breaking self-similarity, and ${\rho\propto r^{-3/2}}$ profiles result.  In no Gaussian formation scenario can both self-similarity and isolation be satisfied, and as a result, the ${\rho\propto r^{-9/4}}$ similarity solution is not physically realized if fluctuations are Gaussian.  We note, however, that an alternative seeding mechanism that generates uncompensated peaks in a relatively smooth background can still yield halos with ${\rho\propto r^{-9/4}}$ profiles, as demonstrated by the collapse of the uniform ellipsoid described in Sec.~\ref{sec:ideal} and of the radial Gaussian overdensity in GABH.

We now give an example of how this correction changes observable UCMH properties.  Some of the most stringent bounds on the primordial power spectrum at small scales come from the nonobservation by Fermi-LAT of gamma-ray sources matching the expected UCMH signal \cite{bringmann2012improved}.  The gamma-ray luminosity $L$ of a UCMH due to WIMP annihilation is proportional to
\begin{equation}\label{fluxint}
L \propto \int_0^R\rho^2(r)r^2\mathrm{d}r,
\end{equation}
where $R$ is an outer boundary of the halo.  Equation~(\ref{fluxint}) diverges for both ${\rho\propto r^{-9/4}}$ and ${\rho\propto r^{-3/2}}$, but in practice, the cusp will flatten out near the center due to annihilation.  We use the estimate
${\rho_\mathrm{max} = m_\chi / [\langle \sigma v \rangle (t-t_\mathrm i)]}$
\cite{berezinsky1992distribution} for the maximum density at time $t$ in a structure formed at $t_i$ owing to annihilation, where $m_\chi$ is the WIMP mass and $\langle\sigma v\rangle$ is its annihilation cross section.  Assuming ${m_\chi = \SI{1}{TeV}}$ and ${\langle\sigma v\rangle = \SI{3e-26}{cm\tothe{3}s\tothe{-1}}}$, we obtain $\rho_\mathrm{max}$ and then evaluate Eq.~(\ref{fluxint}) with the density profile given in Ref.~\cite{bringmann2012improved} for a $\rho\propto r^{-9/4}$ UCMH of scale ${k=\SI{7}{kpc\tothe{-1}}}$.  We then compare\footnote{It is tempting to compare the signals of equal-mass halos, but this is deceptive because while ${\rho\propto r^{-9/4}}$ halos grow with ${M(a)\propto a}$ \cite{bertschinger1985self}, halos with a ${\rho\propto r^{-3}}$ outer profile grow logarithmically or slower \cite{wechsler2002concentrations,lu2006origin}.  Thus, halos of each shape with similar initial masses have wildly different masses today.} this luminosity to that obtained from the fitting form of Fig.~\ref{fig:spike}, and we find that our corrected form reduces the gamma-ray luminosity of the halo by a factor of $200$.

This calculation implies a substantial reduction in UCMH visibility, which would, for instance, raise the upper bound on the number density of UCMHs from the absence of point-source observations (which scales as $L^{-3/2}$) by a factor of $3000$.  However, the loss of constraining power may be counteracted by more sophisticated analyses.  In particular, there is no longer any reason to restrict an analysis to halos forming at ${z\gtrsim 1000}$, and as shown in Fig.~\ref{fig:halo}, the Universe becomes densely populated with halos by ${z=100}$ if density fluctuations are large enough to induce collapse by ${z\simeq 1000}$.  We will develop a formalism for calculating revised constraints in forthcoming work \cite{delos2018}.

\begin{acknowledgments}
The simulations for this work were carried out on the KillDevil computing cluster at the University of North Carolina at Chapel Hill.  The authors would like to thank Erin Conn, Lucas deHart, Josh Horowitz, and Dayton Ellwanger for their valuable assistance in getting this project started on KillDevil.  M.\,S.\,D. and A.\,L.\,E. were partially supported by NSF Grant No. PHY-1417446.  M.\,S.\,D. was also supported by the Bahnson Fund at the University of North Carolina at Chapel Hill.  A.\,P.\,B. contributed to this project while participating in the Computational Astronomy and Physics (CAP) Research Experiences for Undergraduates (REU) program funded by NSF Grant No. OAC-1156614 (PI S. Kannappan).
\end{acknowledgments}

\bibliography{UCMHdensityreferences}

\end{document}